\documentstyle[11pt]{article}

 \csname
@addtoreset\endcsname{equation}{section}

\newcommand{\be}{\begin{equation}}
\newcommand{\ee}{\end{equation}}
\newcommand{\bea}{\begin{eqnarray}}
\newcommand{\eea}{\end{eqnarray}}
\newcommand{\f}{\frac}
\newcommand{\e}{\epsilon}
\newcommand{\al}{\alpha}
\newcommand{\bet}{\beta}

\newcommand{\Ga}{\Gamma}

\newcommand{\vs}[1]{\vspace{#1 mm}}
\newcommand{\hs}[1]{\hspace{#1 mm}}

\begin{document}
\baselineskip=.55cm
\thispagestyle{empty}
\rightline{hep-th 0004199}
\vs{20}

\centerline{\large\bf New Brane Solutions from Killing Spinor Equations}
\vs{15}
\centerline{Ali Kaya\footnote{e-mail: ali@rainbow.tamu.edu}}
\vs{5}
\centerline{Center for Theoretical Physics, Texas A\& M University,}
\centerline{College Station, Texas 77843, USA.}
\vs{10}
\begin{abstract}

In a recent paper, we have pointed out a relation between the Killing spinor and Einstein equations. Using this relation,  new brane solutions of $D=11$ and $D=10$ type IIB supergravity theories are constructed. It is shown that in a brane solution, the  flat world-volume directions, the smeared transverse directions and the sphere located at a fixed radial distance can be replaced with any Lorentzian Ricci flat, Euclidean Ricci flat and Einstein manifolds, respectively. The solution obtained in this fashion is supersymmetric when  the Ricci flat and Einstein manifolds have Killing spinors. We generalize intersecting brane solutions, in which M5, M2 and D3-branes also wrap over the cycles determined by the K\"{a}hler forms of Ricci flat K\"{a}hler manifolds. New, singular, Ricci flat manifolds as (generalized) cones over the U(1) bundles over Ricci flat K\"{a}hler spaces are constructed. These manifolds have covariantly constant spinors and give rise to  new, supersymmetric,  Ricci flat compactifications of non-gauged supergravity theories. We find M2 and D3-brane solutions, which asymptotically approach these singular vacua. 

\end{abstract}
\vs{30}
\pagebreak

\setcounter{page}{1}

\section{Introduction}
\hs{4} Studying brane solutions of low energy supergravity equations is proved to be an important way of searching non-perturbative properties of string/M theories. These  solutions play a crucial role both in the conjectured duality symmetries between seemingly different string/M theories and in the AdS/CFT correspondence. Generically, a brane solution has the interpretation of a $p$-dimensional extended black hole, which can be characterized by few constants like the mass and the charges of the antisymmetric tensor fields. The metric along the world-volume directions have Poincare, and along the transverse directions have rotational invariances. The geometry is asymptotically flat and the number of transverse directions dictates the radial coordinate dependence of the metric functions. By smearing out some transverse directions, solutions with slower fall-off properties can also be constructed (for a review see, for instance, \cite{duff1} \cite{g0})\\

Generalizations of the usual brane solutions have been studied from different points of view. In \cite{duff2}, it has been shown that the transverse 7-sphere of the membrane solution can be replaced with any Einstein manifold. The brane solutions having transverse hyper-K\"{a}hler spaces have been studied in \cite{g2}. In \cite{ak}, a fivebrane solution wrapping on the manifold K3 has been constructed. Brane solutions with Ricci flat world-volumes have been obtained in \cite{perry} \cite{more}. In \cite{curved} \cite{papa}, more general brane solutions with curved world-volume directions have been studied. An intrinsic metric, representing non-trivially embedded D3-branes, has been given in \cite{man}. Brane solutions which are product of an AdS space with an Einstein space have been constructed in \cite{e1} \cite{e2} \cite{e3}. \\    
  
In this paper, we will systematically extend these observations by using a theorem proved in \cite{ak} which states that, under certain conditions, the existence of a Killing spinor implies the  Einstein equations. This enables one to concentrate on the first order Killing spinor equations without worrying about more complicated, second order Einstein equations. The spaces having Killing spinors will play a crucial role in the constructions. \\

The organization of the paper is as follows. In section 2, we review the theorem of \cite{ak}. In sections 3 and 4, we specifically consider M2, M5 and D3-branes. In section 3, we generalize the well known solutions and obtain branes having Ricci flat world-volumes and smeared transverse directions, and non-spherical cross sections. In section 4, using K\"{a}hler forms of Ricci flat K\"{a}hler spaces, we construct intersecting brane solutions which can also be interpreted as wrapping branes  over the cycles determined by K\"{a}hler forms. In section 5, we show that a generic brane background still obeys the field equations when certain directions in the metric are replaced with more general spaces. In section 6, we first construct new, singular, Ricci flat manifolds, which have covariantly constant spinors, as (generalized) cones over U(1) bundles over Ricci flat K\"{a}hler manifolds.  These manifolds  give rise to new supersymmetric compactifications of non-gauged supergravities. In the same section, we find M2 and D3-brane solutions which asymptotically approach these vacua. We conclude with some brief remarks in section 7.

\section{BPS solutions from Killing spinor equations: a theorem} 

\hs{4}It is well known that existence of a covariantly constant spinor on
a Euclidean  manifold implies Ricci flatness. It turns out, this observation has a very useful generalization in the supergravity theory context \cite{ak}. \\

Let us consider a bosonic background ($g_{MN}$,$F_{MNPQ}$) of $D=11$ supergravity theory \cite{cr1} which obeys: 
\be\label{e} R_{MN}=\frac{1}{3}
(F_{M}{}^{PQR} F_{NPQR} -\frac{1}{12} g_{MN}F^{PQRS}F_{PQRS}), 
\ee
\be\label{fe} 
\nabla_{Q}F^{QMNP}=\frac{1}{(24)^{2}}
\epsilon^{MNPA_{1}...A_{8}}F_{A_{1}..A_{4}} F_{A_{5}..A_{8}}.  \ee 
The linearized Rarita-Schwinger equations on this background may be written as:
\be
\label{RS} \Gamma^{MNP}D_{N}\psi_{P} = 0,  
\ee 
where the supercovariant derivative $D_{M}$ is given by
\be
D_{M}=\nabla_{M} +
\frac{1}{144}(\Gamma^{PQRS}{}_{M}-\frac{1}{8}\delta^{P}_{M}\Gamma^{QRS}) 
F_{PQRS}, 
\ee 
and $\nabla_{M}$ is the usual covariant derivative acting on spinors. Let us further consider a linearized spin 3/2 field $\psi_{M}$, which is obtained by the action of supercovariant derivative on an arbitrary spinor $\epsilon$:  
\be\label{F} \psi_{M}=D_{M}\epsilon.
\ee 
Due to the invariance of D=11 supergravity at the linearized fermionic level, this spin 3/2 field solves (\ref{RS}). To verify this claim, we insert (\ref{F}) into (\ref{RS}) and, $using$ $only$ the 4-form field equations, obtain
\be \label{gt}
\Gamma_{M}{}^{NP}D_{N}D_{P}\e=\f{1}{2}(G_{MN} - T_{MN})\Gamma^{N}\epsilon=0,
\ee
where $G_{MN}=R_{MN}-\frac{1}{2} g_{MN}R$ is the Einstein tensor and
\be
T_{MN}=\f{1}{3}(F_{M}{}^{PQR}F_{NPQR}-\f{1}{8}g_{MN}F^{PQRS}F_{PQRS})
\ee
is the energy momentum tensor. In setting (\ref{gt}) to zero, we have used the fact that the background is chosen to obey Einstein equations (\ref{e}).\\

Now consider a background which satisfies the 4-form field equations (\ref{fe}) but $not$ necessarily obeys the Einstein equations. Let us further assume the existence of at least one Killing spinor obeying
\be
D_{M}\epsilon_{0}=0.
\ee 
Following (\ref{gt}),  it is easy to see that 
\be \label{gt2}
(G_{MN} - T_{MN})\Gamma^{N}\epsilon_{0}=0.
\ee
Note that in deriving (\ref{gt}), we have $only$ used the 4-form field equations which  are also assumed to be satisfied by the new background. (\ref{gt}) has been set to zero since that background was chosen to obey Einstein equations. On the other hand, (\ref{gt2}) is satisfied since $\epsilon_{0}$ is the Killing spinor. Therefore, (\ref{gt2}) is still valid, even if the background is not chosen to obey Einstein equations.\\ 

Although (\ref{gt2}) is very close to the Einstein equations, one needs to impose further conditions to proceed due to the Lorentzian signature of the metric. Since $(G_{MN}-T_{MN})$ is symmetric, one can point-wise diagonalize it by elements of the group O(11)\footnote{In claiming this, we assume that there is no topological obstruction to have a continuous map from space-time to the group manifold O(11).} . However, this does not give any new information, since O(11) transformation of O(1,10) gamma matrices are not nice objects. Therefore, one should be able to split space and time directions in (\ref{gt2}). It turns out it is sufficient to assume existence of an orthonormal basis such that $(G_{0i}-T_{0i})=0$, where $0$ is the time and $i$ is the spatial direction. When the indices of (\ref{gt2}) refer to this orthonormal basis, $M=0$ component implies $G_{00}-T_{00}=0$. Along spatial directions, when $M=i$, one can point-wise diagonalize the symmetric matrix $(G_{ij}-T_{ij})$ by O(10) transformations, under which the spatial gamma matrices still satisfy the same Clifford algebra and thus invertible. Therefore, the diagonal entries of $(G_{ij}-T_{ij})$ should also be zero, which then implies Einstein equations $G_{MN}-T_{MN}=0$. Summarizing above considerations, we obtain the following result:\\

(i) if a background is known to solve the 4-form field equations (\ref{fe}),\\

(ii) if there exist an orthonormal basis such that $G_{0i}-T_{0i}=0$; \footnote{Here, we relax the corresponding condition imposed in \cite{ak}. I would like to thank C.N. Pope for the discussions about this point.} \\

(iii) if it is known that there exist a Killing spinor on this  background,\\

then this background, being preserve some fraction of supersymmetries, also solves the Einstein equations of D=11 supergravity.\\

The main advantage of the theorem for applications is that, to find a solution of the second order Einstein equations, one can instead concentrate on the first order Killing spinor equations which are of course easier to solve. One can also convince himself that it is not hard to satisfy conditions (i) and (ii). Indeed, in constructing $p$-brane solutions, one starts with ansatzs having the property (i) and (ii) (see, for instance, \cite{hs} \cite{rg} \cite{t1} \cite{g1} \cite{tsyt}). As we will see for a moment, condition (ii) can easily be satisfied by choosing the background to be static. We finally note that a solution obtained in this fashion already preserves some fraction of supersymmetries.\\  

Although explicitly  proven for $D=11$ supergravity, it is possible to argue that such a theorem can be established  for all supergravities. The linearized supersymmetry invariance of  any supergravity theory requires
\be\label{ep}
\Gamma^{MNP}D_{N}D_{P}\epsilon = 0, 
\ee
where $D_{M}=\nabla_{M}+....$ is the supercovariant derivative, $\epsilon$ is an arbitrary spinor and  bosonic fields refer to a background which obeys equations of motion. In (\ref{ep}),  the derivatives of the metric are expected to appear in the combination of the Einstein tensor\footnote{This is due to the identity $\Gamma_{M}{}^{NP}\nabla_{N}\nabla_{P}\epsilon = 1/2 G_{MN}\Gamma^{N}\epsilon$. }. Therefore, after imposing all but  Einstein equations, (\ref{ep}) should become 
\be\label{yeter}
\Gamma_{M}{}^{NP}D_{N}D_{P}\epsilon = \f{1}{2}(G_{MN}-T_{MN})\Gamma^{N}\e = 0,
\ee
where $T_{MN}$ is the appropriate energy momentum tensor of the theory and in the last step we have used the fact that the background obeys Einstein equations. \\

With this information, it is very easy to prove the theorem; dropping the condition that the background satisfies Einstein equations and further assuming  the existence of a Killing spinor, one obtains (\ref{yeter}) evaluated for the Killing spinor. With condition  (iii), this implies the Einstein equations. Therefore, if we replace (i) with\\

($i^{'}$) if the background satisfies all but Einstein equations,\\

then together with conditions (ii) and (iii) above , the result  of the theorem should apply to all supergravities.\\

To conclude this section, we finally comment on a simple way to satisfy condition (ii). For a static background there exist an orthonormal basis such that $G_{0i}=0$. To see this, using the fact that the background is static, we write the line element as
\be 
ds^{2} = A^{2}(-dt^{2}) + ds_{M}^{2}, 
\ee
where $A$ is a time independent function and $ds_{M}^{2}$ is a line element on the Euclidean manifold $M$. One can easily show that, when expressed in the basis $e^{0}=Adt$ and $e^{i}$ (an orthonormal  basis on $M$),  the Ricci tensor of $ds^{2}$ obeys $R_{0i}=0$. This gives $G_{0i}=0$ since in any orthonormal basis $g_{0i}=0$. On the other hand, if the matter fields are chosen to be static, then $T_{0i}=0$\footnote{This can be regarded as the definition for matter fields to be static.}. Therefore, one immediate way to satisfy condition (ii) is to take background fields to be static. However, one should not rule out existence of interesting non-static cases for which only the combination $(G_{0i}-T_{0i})=0$.

\section{Generalized brane solutions}

\hs{4}As mentioned in the introduction, a generic $p$-brane solution has Poincare invariance on the world-volume and spherical symmetry along transverse directions. In this section, we will focus on the M2, M5 and D3-branes and try to obtain generalizations of the well known  solutions, by using the theorem of section 2. 

\subsection*{Definitions}

\hs{4}Let $L_{d}$, $M_{m}$ and $X_{n}$ be Lorentzian Ricci flat, Euclidean Ricci flat and Einstein manifolds of dimensions $d$, $m$ and $n$, respectively. We will denote the basis one-forms on these spaces as $e^{\mu}$, $e^{a}$ and $e^{\al}$, where the indices $\mu ,\nu ...$ refer to $L_{d}$; $a,b ..$ refer to $M_{m}$ and $\al, \bet.. $ refer to $X_{n}$. We will assume that $L_{d}$ and $M_{m}$ have covariantly constant Killing spinors, 
\be \label{ins0}\nabla_{\mu}\e =0, \ee
\be \nabla_{a}\e = 0,\label{insara} \ee
and $X_{n}$ has Killing spinors obeying
\be\ \label{ins}
\nabla_{\al}\e = -\f{1}{2} \Ga_{\al}\Ga_{r}\e ,
\ee  
where we set the ``inverse radius'' of $X_{n}$ to 1. In these equations, we will not restrict representations of $\Ga$-matrices to be irreducible. Indeed, we will view the Clifford algebras on $L_{d}$, $M_{m}$ and $X_{n}$ as sub-algebras of a bigger Clifford algebra. For instance, in (\ref{ins}), $\Ga_{r}$ can be any element of this bigger algebra which anti-commutes with $\Ga_{\al}$ and squares to identity. We will write the line elements on $L_{d}$, $M_{m}$ and $X_{n}$ as $ds^{2}_{L_{d}}$, $ds^{2}_{M_{m}}$ and  $ds^{2}_{X_{n}}$, respectively. The volume forms will be denoted by $V_{L}$, $V_{M}$ and $V_{X}$. \\

We will use  $'$ to denote differentiation with respect to the argument of the function, and $\sim$ to mean equality up to a constant. The indices in tensor equations will refer to the tangent space. Specifically, $0$ and $i$ will be used for time-like and spatial directions, respectively.

\subsection*{M2-brane}

\hs{4}Let us start from the eleven dimensional membrane. Our aim is to write an ansatz, which satisfies  the conditions (i) and (ii) of the theorem of section 2, and then work out the Killing spinor equations. For the metric and 4-form field we assume 
\bea\label{mxonce}
ds^{2}&=&A^{2}ds_{L_{3}}^{2}+B^{2}dr^{2}+C^{2}ds_{M_{m}}^{2}+D^{2}ds_{X_{n}}^{2},\\ \label{mx}
*F &\sim &V_{M}\wedge V_{X}, 
\eea
where $m+n=7$, $*$ is the Hodge dual corresponding to $ds^{2}$ and the metric functions $A,B,C$ and $D$ are chosen to depend only on the coordinate $r$. We first note that the 4-form field equations are identically satisfied without imposing any condition on the metric functions. It is also easy to see that $G_{0i}=T_{0i}=0$ in the orthonormal basis $E^{\mu}=Ae^{\mu}$, $E^{r}=Bdr$, $E^{a}=Ce^{a}$ and $E^{\al}=De^{\al}$. Therefore, the ansatz obeys the conditions (i) and (ii) of section 2. To find a supersymmetric solution of $D=11$ supergravity, one needs to impose conditions on metric functions which will ensure existence of at least one Killing spinor obeying
\bea\label{kill1}
D_{\mu}\e&=& \nabla_{\mu}\e +\f{A^{'}}{2AB}\Ga_{\mu}{}^{r}\e + \f{q_{e}}{18C^{m}D^{n}}\e_{\nu\rho\sigma}\Ga^{\nu\rho\sigma r}\Ga_{\mu}\e = 0,\\
D_{r}\e &=& \partial_{r}\e + \f{q_{e}}{18C^{m}D^{n}}\e_{\nu\rho\sigma}\Ga^{\nu\rho\sigma}\e =0,\\
D_{a}\e&=& \nabla_{a}\e +\f{C^{'}}{2CB}\Ga_{a}{}^{r}\e + \f{q_{e}}{36C^{m}D^{n}}\e_{\nu\rho\sigma}\Ga^{\nu\rho\sigma r}\Ga_{a}\e = 0,\\ \label{kill4}
D_{\al}\e &=& \nabla_{\al}\e +\f{D^{'}}{2DB}\Ga_{\al}{}^{r}\e + \f{q_{e}}{36C^{m}D^{n}}\e_{\nu\rho\sigma}\Ga^{\nu\rho\sigma r}\Ga_{\al}\e = 0,
\eea
where the indices and spinors refer to the basis one-forms defined above and the electrical charge $q_{e}$ is defined to be the proportionality constant in (\ref{mx}) so that 
\be \label{q}
F_{\mu\nu\rho r}= \f{q_{e}}{C^{m}D^{n}}\e_{\mu\nu\rho}.
\ee
The presence of the functions $C$ and $D$ in (\ref{q}) is due to the fact that $*$ in (\ref{mx})  refers to the line element (\ref{mxonce}). An easy way to solve these equations is to impose (\ref{ins0})-(\ref{ins}) and  
\be\label{pro}
\e_{\mu\nu\rho}\Ga^{\mu\nu\rho}\e = - 6 \e,
\ee
which are of course consistent with each other. Then, the Killing spinor equations (\ref{kill1})-(\ref{kill4}) imply  
\be \label{denk1}
\f{A^{'}}{AB}= \f{2q_{e}}{3C^{m}D^{n}}, \hs{5} \f{C^{'}}{CB}= \f{-q_{e}}{3C^{m}D^{n}},
\hs{5} \f{D^{'}}{DB}= \f{-q_{e}}{3C^{m}D^{n}} + \f{1}{D}.
\ee
Therefore, when (\ref{denk1}) is satisfied, the background has at least one Killing spinor obeying (\ref{ins0}),(\ref{ins}) and (\ref{pro}). Now, by the theorem of section 2, Einstein equations should be satisfied identically. At this point it is instructive to verify this claim by a direct calculation. The Einstein equations can be written as
\be
2\left(\f{A^{'}}{AB}\right)^{2}+ \left(\f{A^{'}}{B}\right)^{'}\f{1}{AB} + m \f{A^{'}C^{'}}{AC}\f{1}{B^{2}} + n \f{A^{'}D^{'}}{AD}\f{1}{B^{2}}=\f{4q_{e}^{2}}{3C^{2m}D^{2n}},
\ee
\be
3\left(\f{A^{'}}{B}\right)^{'}\f{1}{AB} + m \left(\f{C^{'}}{B}\right)^{'}\f{1}{CB} + n \left(\f{D^{'}}{B}\right)^{'}\f{1}{DB} = \f{4q_{e}^{2}}{3C^{2m}D^{2n}},
\ee
\be
(m-1)\left(\f{C^{'}}{CB}\right)^{2}+ \left(\f{C^{'}}{B}\right)^{'}\f{1}{CB} + 3 \f{A^{'}C^{'}}{AC}\f{1}{B^{2}} + n \f{C^{'}D^{'}}{CD}\f{1}{B^{2}}= \f{-2q_{e}^{2}}{3C^{2m}D^{2n}},\\
\ee
\be
-\f{1}{D^{2}}+ (n-1)\left(\f{D^{'}}{DB}\right)^{2} + \left(\f{D^{'}}{B}\right)^{'}\f{1}{DB} + 3 \f{A^{'}D^{'}}{AD}\f{1}{B^{2}} + m \f{C^{'}D^{'}}{CD}\f{1}{B^{2}}=\f{-2q_{e}^{2}}{3C^{2m}D^{2n}}.
\ee
where we have used the fact that $L_{3}$ and $M_{m}$ are Ricci flat and $X_{n}$ is Einstein. It is now straightforward to check that, when the unknown functions $A,B,C$ and $D$ obey (\ref{denk1}) they also solve the Einstein equations, as required by the theorem.\\

There are four independent functions and three differential equations, which may be thought to imply that functions are not constrained enough. However, this is simply a manifestation of reparametrization invariance in coordinate $r$. We want to fix this reparametrization freedom in a way, which will help us in solving  the differential equations. A convenient choice is
\be
C^{m}D^{n-1}=r^{n-1} + M,
\ee
where $M=2q_{e}/(n-1)$. Then, the unknown functions obeying the first order coupled equations (\ref{denk1}) can be solved to give following solution
\bea 
ds^{2}&=&H^{-2/3}ds_{L_{3}}^{2}+ H^{1/3} ( dr^{2} + ds_{M_{m}}^{2}+ r^{2} ds^{2}_{X_{n}}),\\
F_{\mu\nu\rho r}&=&- \f{1}{2} H^{-7/6}(\partial_{r}H) \e_{\mu\nu\rho},
\eea
where  $H=(1+M/r^{n-1})$. 

\subsection*{M5-brane}

\hs{4}For the generalized M5-brane we write the following ansatz,
\bea
ds^{2}&=&A^{2}ds_{L_{6}}^{2}+B^{2}dr^{2}+C^{2}ds_{M_{m}}^{2}+D^{2}ds_{X_{n}}^{2}, \\ \label{fx}
F&\sim& V_{M}\wedge V_{X}, 
\eea
where $m+n=4$ and the metric functions $A,B,C$ and $D$ are chosen to depend only on the coordinate $r$. As for the M2-brane, the 4-from field equations are satisfied identically, and $G_{0i}=T_{0i}=0$ in the orthonormal basis $E^{\mu}=Ae^{\mu}$, $E^{r}=Bdr$, $E^{a}=Ce^{a}$ and $E^{\al}=De^{\al}$. Therefore, the background satisfies conditions (i) and (ii) of section 2. To solve the Killing spinor equations, one can choose $\e$ to obey (\ref{ins0})-(\ref{ins}) and 
\be
\e_{a..b}\e_{\al ..\beta}\Ga^{a..b\al ..\beta r}\e = - m!n!\e .
\ee
which are consistent with each other. One can then check that the Killing spinor equations imply
\be \label{denk2}
\f{A^{'}}{AB}= \f{q_{m}}{3C^{m}D^{n}}, \hs{5} \f{C^{'}}{CB}= \f{-2q_{m}}{3C^{m}D^{n}},\hs{5}
 \f{D^{'}}{DB}= \f{-2q_{e}}{3C^{m}D^{n}} + \f{1}{D},
\ee 
where the magnetic charge $q_{m}$ is defined to be the proportionality constant in (\ref{fx}). Like in the membrane case, a convenient way to fix the $r$-reparametrization invariance is to define $C^{m}D^{n-1}=r^{n-1} +M$, where $M=2q_{m}/(n-1)$. Then, (\ref{denk2}) can be solved to give the following solution
\bea
ds^{2}&=&H^{-1/3}ds_{L_{6}}^{2} + H^{2/3} (dr^{2} + ds_{M_{m}}^{2} + r^{2}ds_{X_{n}}^{2}),\\
F_{a..b\al..\beta}&=& -\f{1}{2} H^{-4/3}(\partial_{r}H)\e_{a..b}\e_{\al ..\beta},
\eea
where $H=(1+M/r^{n-1})$. 

\subsection*{D3-brane}
       
\hs{4}The $D3$-brane solution of IIB supergravity in 10-dimensions has only non-vanishing (anti)-self dual 5-form and the metric. The equations governing dynamics of these fields can be written as \cite{sh1}
\be
R_{MN}=\f{1}{96}F^{PQRS}{}_{M}F_{PQRSN}
\ee
\be
dF=0, *F=-F.
\ee
The Killing spinors on such a background obey
\be
D_{M}\e=\nabla_{M}\e +\f{i}{4\times 480} \Ga^{NP..Q}\Ga_{M}F_{NP..Q}\e=0,
\ee
where $\e$ is a Weyl spinor $\Ga_{(11)}\e=\e$; $\Ga_{(11)}=\Ga^{0}...\Ga^{9}$, $\Ga_{(11)}^{\dagger}=\Ga_{(11)}$ and $\Ga_{(11)}^{2}=I$. For the generalized $D3$-brane we assume the form
\bea 
ds^{2}&=&A^{2}ds_{L_{4}}^{2}+B^{2}dr^{2}+C^{2}ds_{M_{m}}^{2}+D^{2}ds_{X_{n}}^{2}, \\ \label{5form}
F &\sim &(V_{M}\wedge V_{X}) - *(V_{M}\wedge V_{X}),
\eea
where $m+n=5$, and the metric functions depend only on the coordinate $r$. The 5-form field equations are identically satisfied  and  $G_{0i}=T_{0i}=0$ in the basis $E^{\mu}=Ae^{\mu}$, $E^{r}=Bdr$, $E^{a}=Ce^{a}$ and $E^{\al}=De^{\al}$.  Therefore, the ansatz obeys the conditions $(i^{'})$ and (ii) of section 2. On the other hand, if one choose $\e$ to obey (\ref{ins0})-(\ref{ins}) and impose further 
\be
\e_{\mu\nu\rho\sigma}\Ga^{\mu\nu\rho\sigma}\e=24 i  \e,
\ee
the Killing spinor equations imply
\be \label{denk3}
\f{A^{'}}{AB}= \f{q}{4C^{m}D^{n}}, \hs{5} \f{C^{'}}{CB}= \f{-q}{4C^{m}D^{n}},\hs{5}
 \f{D^{'}}{DB}= \f{-q}{4C^{m}D^{n}} + \f{1}{D},
\ee 
where the dyonic charge $q$ is defined to be the proportionality constant in (\ref{5form}). Therefore, when $A,B,C$ and $D$ obey (\ref{denk3}), all conditions of the theorem are satisfied and the background should obey Einstein equations. Fixing $r$-reparametrization invariance by $C^{m}D^{n-1}=r^{n-1} +M$, we obtain the following solution,
\bea
ds^{2}&=&H^{-1/2}ds_{L_{4}}^{2} + H^{1/2}(dr^{2}  + ds_{M_{m}}^{2} + r^{2}ds_{X_{n}}^{2}), \\
F_{\mu\nu\rho\sigma r} &=& -\f{1}{2}H^{-5/4}(\partial_{r}H)\e_{\mu\nu\rho\sigma},
\eea
where $M=q/(n-1)$ and $H=(1+M/r^{n-1})$.

\subsection*{Interpretation and special cases}

\hs{4}It is clear  from the structure of the antisymmetric tensor fields and Killing spinor projections that, in all solutions, $L_{d}$ represents the curved world-volumes of the branes. The dependence of the metric functions on the radial coordinate $r$ implies that $M_{m}$ and $X_{n}$ correspond to the smeared and actual transverse directions, respectively. \\

The solutions obtained so far have a very similar structure with the well known brane solutions. Indeed, when $X_{n}$ is chosen to be the $n$-sphere, $(r,X_{n})$ space becomes flat. In this case, $H$ can be generalized to be any harmonic function on this flat space. Choosing, furthermore, $L_{d}$ and $M_{m}$ to be flat gives the well known M2, M5 and D3-brane solutions which have $m$ smeared transverse directions. Therefore, one can think of the new solutions as the branes having curved world-volumes and smeared transverse directions, and non-spherical cross sections. \\

For $m=0$, i.e. when $M_{m}$ is empty, one obtains the solutions of \cite{e1} \cite{e2} \cite{e3}, which can thus be viewed to be the members of a more general family found in this paper. For $n=1$, reparametrization fixing conditions should clearly be modified. A convinient choice is to impose $D=rC$, which then gives the same solutions above with $H=q \log r + const.$\\

The number of Killing spinors on $L_{d}$, $M_{m}$ and $X_{n}$ determines the number of unbroken supersymmetries. Finally, it is also worth to mention that, field equations are still satisfied even when  $L_{d}$, $M_{m}$ and $X_{n}$ have no Killing spinors. As we will show in section 5, this is not a coincidence, and indeed there is another simple way of generating new solutions.

\section{Generalized intersections}

\hs{4}To generalize intersecting M2, M5 and D3-brane solutions, we will make use of Ricci flat K\"{a}hler spaces. In this section, we will still use the definitions of section 3, but furthermore  assume that $M_{m}$ has a covariantly constant complex structure obeying
\be
J_{a}{}^{b}J_{b}{}^{c}=-\delta_{a}{}^{c},
\ee
\be 
\nabla_{c}J_{ab}=0.
\ee 
This implies that $M_{m}$ is a Ricci flat K\"{a}hler space, $m$ is an even integer and $J$ is the K\"{a}hler two-form. Our basic strategy will still be the same; we will write an ansatz obeying conditions (i) and (ii) of section 2, and then work out the Killing spinor equations.

\subsection*{M2-brane intersections}

\hs{4}We start with the following ansatz
\bea 
ds^{2}&=&A^{2}(-dt^{2})+B^{2}dr^{2}+C^{2}ds_{M_{m}}^{2}+D^{2}ds_{X_{n}}^{2}, \\\label{intm2}
* F&\sim &(*_{M}J)\wedge V_{X},
\eea
where $n+m=9$ with $n=1,3,5,7$, $*_{M}$ is the Hodge dual on the manifold $M_{m}$ and the metric functions are assumed to depend only on $r$. For $n=7$, the ansatz becomes a special case of the generalized M2-brane ansatz of section 3, and thus we will mainly consider $n=1,3,5$ cases. The 4-form field equations are identically satisfied since $J$ is both closed and co-closed on $M_{m}$. Furthermore, $G_{0i}=T_{0i}=0$ in the basis $E^{0}=Adt$, $E^{r}=Bdr$, $E^{a}=Ce^{a}$ and $E^{\al}=De^{\al}$. Therefore, the conditions (i) and (ii) of section 2 are satisfied. To solve the Killing spinor equations, $\e$ can consistently be chosen to obey (\ref{ins0}), (\ref{insara}) and 
\be\label{pro2}
\partial_{t}\e=0,\hs{8} \Ga^{0}\e=i\e,\hs{8} J_{ab}\Ga^{b}\e=i\Ga_{a}\e,
\ee
which implies
\be \label{denkintm2}
\f{A^{'}}{AB}= \f{(9-n)q_{e}}{3C^{7-n}D^{n}}, \hs{5} \f{C^{'}}{CB}= \f{(n-3)q_{e}}{6C^{7-n}D^{n}}, \hs{5}\f{D^{'}}{DB}= \f{-(9-n)q_{e}}{6C^{7-n}D^{n}} + \f{1}{D},
\ee 
where $q_{e}$ is the proportionality constant in (\ref{intm2}). When $n\not = 1$, one can fix $r$-reparametrization invariance by imposing  $C^{m-2}D^{n-1}=r^{n-1}+M$ and this gives the following solution
\bea
ds^{2}&=& H^{\f{(n-9)}{3}}(-dt^{2}) + H^{\f{(3-n)}{6}}ds_{M_{m}}^{2} + H^{\f{(9-n)}{6}}(dr^{2}+r^{2}ds_{X_{n}}^{2}),\\
F_{0rab}&=& -\f{1}{2} H^{\f{(n-21)}{12}}(\partial_{r}H) J_{ab},
\eea
where $M=2q_{e}/(n-1)$ and $H=(1+M/r^{n-1})$. For $n=1$, one should modify reparametrization fixing condition. A convinient choice is to demand $D=rC^{4}$, which then gives the above solution with $H=2q_{e} \log r + const.$ As it will be made more clear when we discuss the special cases, for $n=5,3,1$, the solutions describe two, three and four M2-brane intersections over a line, respectively. The structure of the background 4-form field implies that membranes also wrap over the 2-cycle dual to K\"{a}hler form of the space $M_{m}$. \\

\subsection*{M5-brane intersections}

\hs{4}The discussion for intersecting M5-branes is not very different from M2-brane intersections. Considering the fact that the 4-form field should give rise to magnetic type of charges, we start with the following ansatz
\bea 
ds^{2}&=&A^{2}ds_{L_{d}}^{2}+B^{2}dr^{2}+C^{2}ds_{M_{m}}^{2}+D^{2}ds_{X_{2}}^{2}, \\ \label{intm5}
F &\sim &J\wedge V_{X},
\eea
where $m+d=8$ with $m=2,4,6$ and the metric functions are assumed to depend only on $r$. For $m=2,$ this reduces to a special case of the generalized M5-brane ansatz of section 3. The 4-form field equations are identically satisfied since $J$ is both closed and co-closed on $M_{m}$. On the other hand $G_{0i}=T_{0i}=0$ in the basis $E^{0}=Adt$, $E^{r}=Bdr$, $E^{a}=Ce^{a}$, $E^{\al}=De^{\al}$ and therefore, the conditions (i) and (ii) of section 2 are satisfied. To solve the Killing spinor equations $\e$ can consistently be chosen to obey (\ref{ins0})-(\ref{ins}) and 
\be
\e_{\al\beta}\Ga^{\al\beta r}\e=2i\e, \hs{7} J_{ab}\Ga^{b}\e=i\Ga_{a}\e,
\ee
which implies 
\be \label{denkintm5}
\f{A^{'}}{AB}= \f{mq_{m}}{6C^{2}D^{2}}, \hs{5} \f{C^{'}}{CB}= \f{(m-6)q_{m}}{6C^{2}D^{2}}, \hs{5}\f{D^{'}}{DB}= \f{-mq_{m}}{3C^{2}D^{2}} + \f{1}{D},
\ee 
where $q_{m}$ is the proportionality constant in (\ref{intm2}). To fix the $r$-reparametrization invariance we impose $C^{2}D=r +M $, where $M=2q_{m}$. Then, (\ref{denkintm5}) can be solved to give
\bea
ds^{2}&= & H^{-m/6}ds_{L_{8-m}}^{2} + H^{\f{(6-m)}{6}}ds_{M_{m}}^{2} +  H^{m/3}( dr^{2} + r^{2}ds_{X_{2}}^{2}),\\
F_{ab\al\beta}&=& -\f{1}{2} H^{-\f{(m+6)}{6}}(\partial_{r}H) J_{ab}\e_{\al\beta},
\eea
where $H=(1+M/r)$. As indicated above, for $m=2$ the solution becomes one of the M5-brane solution of section 3. On the other hand, for $m=4$ and $m=6$, the solutions describe two and three M5-branes intersecting over a three-brane and a string, respectively. The structure of the background 4-form field implies that M5-branes also wrap over the $(m-2)$-cycle dual to $*_{M}J$.

\subsection*{D3-brane intersections}

\hs{4}Let us start by discussing possible ansatzs for the (anti)-self dual 5-form involving the K\"{a}hler two-form of $M_{m}$. The first obvious choice is to assume $F\sim J\wedge V_{X_{3}}$. Among the possible cases, $m=2$ corresponds to the smeared D3-brane of section 3 and $m=6$ is not allowed since this does not leave any room for a time-like direction. One may also try to write an ansatz involving $*_{M}J$ such as $F\sim *_{M}J\wedge V_{X_{n}}$. For $m=4$, this becomes equal to the choice $F\sim J\wedge V_{X_{3}}$ and for $m=6$, $X_{n}$ space becomes one-dimensional. On the other hand, $m=2$ case turns out to be the same with the D3-brane ansatz of section 3. Summarizing, as a non-trivial ansatz representing intersecting D3-branes, one can write   
\bea 
ds^{2}&=&A^{2}ds_{L_{2}}^{2}+B^{2}dr^{2}+C^{2}ds_{M_{4}}^{2}+D^{2}ds_{X_{3}}^{2}, \\ \label{intd3}
F &\sim &(J\wedge V_{X})- *(J\wedge V_{X}),
\eea
where the metric functions are assumed to depend only on $r$. The 5-form field equations are identically satisfied and one can check that the background obeys condition (ii) of the section 2. To determine unknown functions, we solve the Killing spinor equations by imposing (\ref{ins0})-(\ref{ins}) and 
\be
\e_{\mu\nu}\Ga^{\mu\nu}\e=2\e, \hs{7} J_{ab}\Ga^{b}\e=i\Ga_{a}\e,
\ee
which implies 
\be \label{denkintD3}
\f{A^{'}}{AB}= \f{q}{2C^{2}D^{3}}, \hs{5} \f{C^{'}}{CB}= 0, \hs{5} \f{D^{'}}{DB}= \f{-q}{2C^{2}D^{3}} + \f{1}{D},
\ee 
where $q$ is the proportionality constant in (\ref{intd3}). Reparametrization invariance can be fixed by demanding $C^{2}D^{2}=r^{2}+M$, where $M=q/2$. Then, (\ref{denkintD3}) can be solved to obtain the solution 
\bea
ds^{2}&=& H^{-1} ds_{L_{2}}^{2} +  ds_{M_{4}}^{2} + H(dr^{2}+r^{2}ds_{X_{3}}^{2}), \\
F_{\mu\nu abr}&=& -\f{1}{2} H^{-3/2}(\partial_{r}H) J_{ab}\e_{\mu\nu},
\eea
where $H=(1+M/r^{2})$, which describes two D3-branes intersecting over a string and wrapping over the cycle dual to $J$. 

\subsection*{Interpretation and special cases}

\hs{4}To be able to interpret these solutions properly, let us choose $X_{n}$ to be the $n$-sphere. In this case, one can introduce  Cartesian coordinates to span  $(r,X_{n})$ space and $H$ can be generalized to be any harmonic function of these coordinates. When $L_{d}$ and $M_{m}$ are also chosen to be flat, the solutions become the well known intersecting brane solutions in which $all$ harmonic functions are $equal$\footnote{In the next section, we will illustrate with an example how to generalize intersecting brane solutions when harmonic functions are not equal.}. \\

Comparing with this special case, it is easy to argue that $L_{d}$, $M_{m}$ and $X_{n}$ correspond to common tangent, relative transverse  and overall transverse directions.  As mentioned earlier, the branes also wrap over the cycles dual to the K\"{a}hler two-form $J$ or its Hodge dual $*_{M}J$. For a given solution, this cycle can be written as a union of $m/2$ different submanifolds. To see this we note that, in an orthonormal basis, $J$ can be written as $J=e^{1}\wedge e^{2}+..+e^{m-1}\wedge e^{m}$. Therefore, the cycle dual to $J$ or $*_{M}J$ becomes the sum of $m/2$ submanifolds dual to two-forms $e^{1}\wedge e^{2}$,..,$e^{m-1}\wedge e^{m}$ or $(m-2)$-forms $*_{M}(e^{1}\wedge e^{2})$, ..,$*_{M}(e^{m-1}\wedge e^{m})$, respectively. On the other hand, the Killing spinor projections imply that there are $m/2$ intersecting branes each wrapping over one of these submanifolds. We note that this interpretation is consistent with the special case where $M_{m}$ is flat.\\

When $X_{n}$ is different from the $n$-sphere, one obtains new solutions which have not been encountered before. These solutions can be viewed to be the intersecting brane counterparts of the brane solutions constructed in \cite{e1}\cite{e2}\cite{e3}, which have non-spherical transverse spaces. \\

In all cases the number of unbroken supersymmetries depend on the number of Killing spinors on $L_{d}$, $M_{m}$ and $X_{n}$. Like for the generalized brane solutions of the previous section, one can check that the field equations are still satisfied, even when the spaces $L_{d}$, $M_{m}$ and $X_{n}$ have no Killing spinors. 

\section{A general argument}

\hs{4}In the last two sections, we have obtained supersymmetric, generalized brane solutions by writing suitable ansatzs and working out Killing spinor equations. As pointed earlier, the field equations are still satisfied, even when the manifolds $L_{d}$, $M_{m}$ and $X_{n}$ have no Killing spinors. This suggests existence of a general relation which may hold at the level of field equations. \\

Let us consider a brane solution which has a metric of the form,
\be\label{met}
ds^{2}= A^{2}ds_{M}^{2} + ds^{2}_{N},
\ee
where $ds_{M}^{2}$ and $ds^{2}_{N}$ are line elements of $m$,$n$ dimensional manifolds $M$ and $N$, respectively, and $A$ is a function on $N$. We assume that the anti-symmetric tensor fields of the solution are of the form $F\sim V_{M}\wedge ...$ or $F\sim ...$, where $V_{M}$ is the volume form of $M$ and  the dotted terms depend only on $N$. Furthermore, we consider the cases in which scalar fields are independent of $M$ and the anti-symmetric tensor field equations reduce to $dF=0$ and $d*F=0$ \footnote{It seems one can relax the last assumption, but here, for simplicity, we do not consider more general cases.}. \\

Our claim is that, for such a brane solution, the field equations are still satisfied when $M$ is replaced with a different manifold $\tilde{M}$, provided the Ricci tensors of both manifolds have the same form. For instance, if $M$ is flat, $\tilde{M}$ can be any manifold which is Ricci flat. Or, if $M$ is a sphere, $\tilde{M}$ can be any Einstein manifold.\\

To prove this, let $Ae^{a}$ and $e^{\al}$ be a basis for the tangent space, where $e^{a}$ and $e^{\al}$  are the basis-one forms of $M$ and $N$. We calculate the Ricci tensor  of (\ref{met}) as
\bea\label{Ricci}
R_{a}{}^{b}&=&\f{1}{A^{2}}R_{(M)a}{}^{b} - (m-1)\f{A_{\al}A^{\al}}{A^{2}}\delta_{a}{}^{b} - \f{F_{\al}{}^{\al}}{A}\delta_{a}{}^{b},\\ \label{Ricci2}
R_{\al}{}^{\beta}&=& R_{(N)\al}{}^{\beta} - n \f{F_{\al}{}^{\beta}}{A}, 
\eea
where the indices refer to the tangent space and $R_{(M)ab}$, $R_{(N)\al\beta}$ are the Ricci tensors of $M$ and $N$, respectively. The tensor quantities $A_{\al}$ and $F_{\al\beta}$ are defined by the relations
\be
dA=A_{\al}e^{\al},
\ee
\be
dA_{\al}=F_{\al\beta}e^{\beta}.
\ee
We note that the Ricci tensor of (\ref{met}) depends only on the dimension $m$ and Ricci tensor $R_{(M)ab}$ of the manifold $M$. \\

Let us now analyze how the field equations may change when one replaces $M$ with $\tilde{M}$. We first note that $F$ is closed or co-closed irrespective of the choice of $M$, thus the form equations are still satisfied. The possible terms in scalar and Einstein equations have at most second order covariant derivatives of scalars. One can check that when $M$ is replaced with $\tilde{M}$, only $\omega^{a}{}_{bc}$ components of the spin connection  may change. This ensures that the second order covariant derivatives of scalars remain the same, since they are assumed to be independent of $M$. Referring to the tangent space, it is easy to show that any tensor field which is constructed from $F$, the terms containing up to second order covariant derivatives of scalars  and the metric have the same form.  By (\ref{Ricci}) and (\ref{Ricci2}), the last statement ensures that both scalar and Einstein equations are still satisfied.\\

In some cases, $M$ can be a K\"{a}hler space and the anti-symmetric tensor fields may be of the form $F\sim J\wedge ..$, where $J$ is the K\"{a}hler two-form. In such cases, $\tilde{M}$ should also be chosen to be K\"{a}hler and $J$ should be replaced with the K\"{a}hler two-form of $\tilde{M}$. One can easily argue that the field equations are not affected from these replacements.\\

Let us now present some examples which enables one to recover some solutions obtained in the literature and in the previous sections. Consider, for instance, the well known smeared M2-brane solution,
\bea 
ds^{2}&=&H^{-2/3}(-dt^{2}+dx_{1}^{2}+dx_{2}^{2})+ H^{1/3} ( dy_{1}^{2}+ ..+dy_{m}^{2}+ dy_{m+1}^{2}+..+dy_{10}^{2}),\\
F_{\mu\nu\rho \alpha}&=&- \f{1}{2} H^{-7/6}(\partial_{\alpha}H) \e_{\mu\nu\rho},
\eea 
where $H$ is harmonic on $(y_{m+1},..,y_{10})$. Referring to the above discussion, it is easy to see that the world-volume directions $(t,x_{1},x_{2})$ and the smeared transverse directions $(y_{1},..,y_{m})$ can be replaced with more general Ricci flat manifolds. In this way, one can obtain  
\bea 
ds^{2}&=&H^{-2/3}ds_{L_{3}}^{2}+ H^{1/3} (ds_{M_{m}}^{2}+ dy_{m+1}^{2}+..+dy_{10}^{2}),\\
F_{\mu\nu\rho \alpha}&=&- \f{1}{2} H^{-7/6}(\partial_{\alpha}H) \e_{\mu\nu\rho},
\eea
which becomes one of the generalized M2-brane solutions constructed in section 3.\\

In IIB theory, one can start from the single centered D3-brane solution
\bea
ds^{2}&=&H^{-1/2}(-dt^{2}+dx_{1}^{2}+..+dx_{3}^{2}) + H^{1/2}(dr^{2} +  r^{2}d\Omega_{5}^{2}), \\
F_{\mu\nu\rho\sigma r} &=& -\f{1}{2}H^{-5/4}(\partial_{r}H)\e_{\mu\nu\rho\sigma},
\eea
where $d\Omega_{5}^{2}$ is the line element on the $5$-sphere and  $H=(1+M/r^{4})$. The metric and the 5-form field are of the form discussed above with the flat world-volume $(t,x_{1},..,x_{3})$ and the transverse $5$-sphere play the role of manifold $M$ in (\ref{met}). By the claim we have just proved, one can generalize the well known solution as 
\bea
ds^{2}&=&H^{-1/2}ds^{2}_{L_{4}} + H^{1/2}(dr^{2} +  r^{2}ds^{2}_{X_{5}}), \\
F_{\mu\nu\rho\sigma r} &=& -\f{1}{2}H^{-5/4}(\partial_{r}H)\e_{\mu\nu\rho\sigma},
\eea
which corresponds to the solution obtained in \cite{e1} when $L_{4}$ is flat. \\

Let us finally consider an example to illustrate how the intersecting solutions can be generalized when the harmonic functions are not chosen to be equal. The background representing two M5-branes intersecting over a three-brane is given by \cite{g1}
\bea \nonumber
ds^{2}&=&(H_{1}H_{2})^{-1/2}(-dt^{2}+ dx_{1}^{2}+dx_{2}^{2}+dx_{3}^{2}) + H_{1}^{-1/3}H_{2}^{2/3}(dx_{4}^{2}+dx_{5}^{2})\\ 
& + &H_{1}^{2/3}H_{2}^{-1/3}(dx_{6}^{2}+dx_{7}^{2}) +(H_{1}H_{2})^{2/3} (dx_{8}^{2}+..+dx_{10}^{2}),\\ \label{intex}
F &\sim & dx_{4} \wedge dx_{5}\wedge * dH_{2} + dx_{6}\wedge dx_{7}\wedge *dH_{1}, \label{intex2}
\eea
where, $H_{1}$ and $H_{2}$ are harmonic and $*$ is the Hodge dual on $(x_{8},..,x_{10})$ space. Here, it is easy to see that, $(t,x_{1},x_{2},x_{3})$, $(x_{4},x_{5})$ and $(x_{6},x_{7})$ spaces play the role of manifold $M$ in (\ref{met}). Since all these spaces are flat, one can replace them with arbitrary Ricci flat manifolds $L_{4}$, $M_{2}$ and $\tilde{M}_{2}$ to obtain \footnote{In two-dimensions, Ricci flatness implies flatness. Here, we would like to emphasize that for field equations to be satisfied Ricci flatness  of these manifolds is sufficient.} 
\bea\nonumber
ds^{2}&=&(H_{1}H_{2})^{-1/3}ds_{L_{4}}^{2} + H_{1}^{-1/3}H_{2}^{2/3}ds^{2}_{M_{2}} + H_{1}^{2/3}H_{2}^{-1/3}ds^{2}_{\tilde{M}_{2}}\\ 
 &+&(H_{1}H_{2})^{2/3} (dx_{5}^{2}+..+dx_{10}^{2}),\\
F &\sim& V_{M_{2}}\wedge *dH_{2} + V_{\tilde{M}_{2}}\wedge *dH_{1}.
\eea
When $H_{1}$=$H_{2}=H$, the 4-form field (\ref{intex2}) can be written as 
\be
F\sim J\wedge *dH,
\ee
where $J$ is the complex structure of the flat space $(x_{4},x_{5},x_{6},x_{7})$. In this case, these four flat directions can be replaced with any 4-dimensional Ricci flat K\"{a}hler space, which corresponds to one class of intersections found in section 4. \\

As can be inferred from these examples, in a brane solution the flat world-volume and smeared transverse directions, and the transverse sphere at a fixed radial distance can be replaced with more general Ricci flat and Einstein manifolds. In intersecting brane solutions the common tangent and relative transverse directions have this property. We note that, these replacements can be done at the level of field equations and supersymmetry of the new solution obtained in this way is not manifest. 

\section{Solutions from $U(1)$ bundles over Ricci flat K\"{a}hler spaces}

\hs{4}In the brane solutions obtained in section 3 the warping factors are the same for the transverse and world-volume directions. In this section, by using  the theorem of section 2,  we will take a step in obtaining solutions which fail to have this property, at least along the transverse directions.  \\

We first construct singular, Ricci flat manifolds having a cone-like structure over U(1) bundles over Ricci flat K\"{a}hler spaces, which give rise to new supersymmetric vacua for non-gauged supergravities. Since the conformal factors multiplying U(1) fibers and the base spaces turn out to be neither equal to each other nor equal to the square of the coordinate parametrizing the cone, we name these spaces as (generalized) cones.\\

As we will see in a moment, there are two singularities associated with the cone, since at the origin the base space and at infinity the U(1) fibers shrink to zero size. Also, in viewing the solutions as Ricci flat compactifications, unlike the conventional Kaluza-Klein picture, there is no natural way of assuming the internal spaces to be small compared to the space-time. \\

At the end of this section, we will construct brane solutions which asymptotically approach these singular compactifications and preserve half of the available supersymmtries. This is in fact not surprising, if one assumes stability of the vacua, since the fundamental branes underlying the supergravity theories reveals themselves as supersymmetric soliton solutions. 

\subsection*{A singular Ricci flat space}

\hs{4}It is well known that the cone over a manifold $X$
\be\label{cone}
ds^{2}=dr^{2}+r^{2}ds_{X}^{2}
\ee
is Ricci flat if and only if $X$ is Einstein. In this section, we would like to construct Ricci flat manifolds having a cone-like structure over Ricci flat K\"{a}hler spaces. It is clear that one should modify (\ref{cone}) in a non-trivial way. Let us consider a metric of the form
\be\label{me}
ds^{2}=B^{2}dr^{2}+C^{2}ds^{2}_{M_{m}}+D^{2}(d\tau - {\cal A})^{2},
\ee
where $M_{m}$ is an $m$-dimensional Ricci flat K\"{a}hler space, ${\cal A}$ is the one-form potential for the K\"{a}hler two-form so that $d{\cal A}=J$, $\tau$ is a periodic coordinate and the metric functions are assumed to depend only on $r$. It is clear that we have a fiber bundle structure and $\tau$ is the coordinate on U(1) fibers. We would like to determine  $B$,$C$ and $D$ which give rise to a Ricci flat space. Instead of calculating Ricci tensor and solving second order, coupled differential equations we demand existence of a covariantly constant spinor which, by considerations of section 2, will imply Ricci flatness. In the tangent space basis $E^{r}=Bdr$, $E^{a}=Ce^{a}$ and $E^{D}=D(d\tau -{\cal A})$, a covariantly constant spinor  obeys
\bea\nonumber
D\e &=&d\e +\f{1}{4}(\omega_{ab}+\f{D}{2C^{2}}J_{ab}E^{D})\Gamma^{ab}\e - \f{1}{4}\f{D}{C^{2}}J_{ab}E^{b}\Gamma^{Da}\e\\
&+& \f{1}{2}\f{C^{'}}{CB}E_{a}\Gamma^{ar}\e +\f{1}{2}\f{D^{'}}{DB}E_{D}\Gamma^{Dr}\e=0,
\eea
where $\omega_{ab}$ is the spin connection on $M_{m}$, and we have used the differential form notation. Conistently imposing 
\be
\partial_{r}\e =\partial_{\tau}\e=0,\hs{10}d\e+\f{1}{4}\omega_{ab}\Gamma^{ab}\e=0, 
\ee
\be
J_{ab}\Gamma^{b}\e=i\Gamma_{a}\e, \hs{5} \Gamma^{Dr}\e =i\e,
\ee
one obtains
\be \label{eger}
\f{C^{'}}{CB}=\f{D}{2C^{2}}, \hs{5}\f{D^{'}}{DB}=-\f{mD}{4C^{2}}.
\ee
Therefore, when the metric functions $B$, $C$ and $D$ obey (\ref{eger}), the Ricci tensor of  (\ref{me}) vanishes. Fixing  $r$-reparametrization invariance by $B=4/(m+4)$, (\ref{eger}) can be solved to give the following Ricci flat metric
\be
ds^{2}=\f{16}{(m+4)^{2}}dr^{2}+r^{4/(m+4)}ds^{2}_{M_{m}}+r^{-2m/(4+m)}(d\tau -{\cal A})^{2},
\ee
which we name as the space $M_{C}$. Topologically, $M_{C}$ is a product of a line parametrized by $r$ and a U(1) bundle over a Ricci flat K\"{a}hler space. The metric is singular, since at the origin, when $r=0$, the base space and at infinity, as $r\to\infty$, the U(1) fibers shrink to zero size. On the other hand, $M_{C}$ have covariantly constant spinors, and thus gives rise to supersymmetric compactifications of non-gauged supergravities. In these compactifications, one can view the coordinate $r$ as a radial coordinate in the uncompactified space-time. Therefore, the U(1) bundle over $M_{m}$ plays the role of internal space. As can be inferred from the structure of the metric, there is no natural way of assuming the internal directions  to be ``small''. Thus these compactifications are very different from the conventional Kaluza-Klein ones. Although the metric is singular, the existence of unbroken supersymmetries is a sign for stability. We now construct M2 and D3-brane solutions which belong to these singular vacua.

\subsection*{M2 and D3-branes  on $M_{C}$ compactifications}

\hs{4}In searching brane solutions which asymptotically belong to above Ricci flat compactifications, we consider the cases in which $M_{C}$ plays the role of the total transverse space. Since $M_{C}$ is an even dimensional manifold, this leaves the possibility of an even dimensional brane in $D=11$, and an odd dimensional brane in $D=10$. We think of the coordinate $r$ as the radial coordinate in the transverse space. \\

M2-brane is the only even dimensional brane in $D=11$, and this suggests the following ansatz
\bea 
ds^{2}&=&A^{2}ds_{L_{3}}^{2}+B^{2}dr^{2}+C^{2}ds^{2}_{M_{6}}+D^{2}(d\tau - {\cal A})^{2},\\ \label{sssss}     
 *F &\sim &V_{M}\wedge(d\tau - {\cal A}),\label{ff2}
\eea
where all metric functions are assumed to depend on $r$. We remind the reader that we are still using the definitions of section 3. One can check that, the 4-form field equations are identically satisfied, and $G_{0i}=T_{0i}=0$ in the basis $E^{\mu}=Ae^{\mu}$, $E^{r}=Bdr$, $E^{a}=Ce^{a}$ and $E^{D}=D(d\tau - {\cal A})$. Therefore, the conditions (i) and (ii) of section 2 are satisfied. By imposing \footnote{The covariant derivatives in (\ref{cov1}) refer to the spaces $L_{3}$ and $M_{6}$.}
\be\label{cov1}
\partial_{\tau}\e=0,\hs{5}\nabla_{\mu}\e=0,\hs{5} \nabla_{a}\e = 0,
\ee
and 
\be
\e_{\mu\nu\rho}\Gamma^{\mu\nu\rho}\e=6\e ,\hs{5}J_{ab}\Gamma^{b}\e=i\Gamma_{a}\e, \hs{5} \Gamma^{Dr}\e =i\e,
\ee
the Killing spinor equations imply
\be\label{mcmet}
\f{A^{'}}{AB}=\f{2q_{e}}{3C^{6}D},\hs{5}\f{C^{'}}{CB}=\f{D}{2C^{2}}-\f{q_{e}}{3C^{6}D},\hs{5}
\f{D^{'}}{DB}=-\f{3D}{2C^{2}}-\f{q_{e}}{3C^{6}D},
\ee
where $q_{e}$ is the proportionality constant in (\ref{ff2}). Since the conditions imposed on the Killing spinor are consistent with each other, when (\ref{mcmet}) is satisfied, the background should obey the Einstein equations by the theorem of section 2. We fix the $r$-reparametrization invariance by imposing $C^{4}D^{2}=r$. Using this condition, (\ref{mcmet}) can be solved to give the following solution
\bea \nonumber
ds^{2} &=& \left(1-\f{M}{r}\right)^{2/3} ds_{L_{3}}^{2}+ \f{1}{r^{7}}\left(1-\f{M}{r}\right)^{-22/3}dr^{2} + \f{1}{r}\left(1-\f{M}{r}\right)^{-4/3} ds_{M_{6}}^{2}\\
 & + & r^{3}\left(1-\f{M}{r}\right)^{8/3}(d\tau - {\cal A})^{2},\\
F_{\mu\nu\rho r} & = & \f{M}{2}r^{3/2}\left(1-\f{M}{r}\right)^{8/3}\e_{\mu\nu\rho},
\eea
where $M=2q_{e}$. \\

It is clear that, in the solution, $L_{3}$ represents the world-volume of the M2-brane and $r$ is a radial coordinate along transverse directions. Unlike all solutions obtained before, there are three different \footnote{Indeed, by a coordinate change, the warping factor along the coordinate $r$ can be made equal, for instnce, to the warping factor along U(1) fiber.} warping factors multiplying the transverse directions, which have also a non-trivial topological structure due to the presence of U(1) bundle. There is a horizon located at $r=M$ and asymptotically, as $r \to \infty$,  the solution becomes $L_{3}\times M_{C}$, which can be seen by a coordinate change $d\tilde{r}\sim r^{-7/2}dr$, near infinity. The solution is a half supersymmetry preserving state of the vacuum $L_{3}\times M_{C}$ since the presence of the M2-brane brakes only half of the available supersymmetries.\\

As an example in $D=10$, we consider the D3-brane of IIB theory and start with the following ansatz
\bea 
ds^{2}&=&A^{2}ds_{L_{4}}^{2}+B^{2}dr^{2}+C^{2}ds^{2}_{M_{4}}+D^{2}(d\tau - {\cal A})^{2},\\
*F &\sim &V_{M}\wedge(d\tau - {\cal A}) - * [V_{M}\wedge(d\tau - {\cal A})] ,\label{ffd3}
\eea
where, as usual, we assume that $A$, $B$, $C$ and $D$ depend only on $r$. One can check that the ansatz obeys the condition $(i^{'})$ and (ii) of section 2, and therefore to obtain a supersymmetric solution one needs to work out Killing spinor equations. Imposing \footnote{The covariant derivatives in (\ref{cov2}) refer to the spaces $L_{4}$ and $M_{4}$.}
\be\label{cov2}
\partial_{\tau}\e=0,\hs{5}\nabla_{\mu}\e=0,\hs{5} \nabla_{a}\e = 0,
\ee
and 
\be
\e_{\mu\nu\rho\sigma}\Gamma^{\mu\nu\rho\sigma}\e=-24i\e ,\hs{5}J_{ab}\Gamma^{b}\e=i\Gamma_{a}\e, \hs{5} \Gamma^{Dr}\e =i\e,
\ee
the Killing spinor equations imply
\be
\f{A^{'}}{AB}=\f{q}{4C^{4}D},\hs{5}\f{C^{'}}{CB}=\f{D}{2C^{2}}-\f{q}{4C^{4}D},\hs{5}
\f{D^{'}}{DB}=-\f{D}{C^{2}}-\f{q}{4C^{4}D},
\ee
where the dyonic charge $q$ is defined to be the proportionality constant in (\ref{ffd3}). Fixing $r$-reparametrization invariance by $C^{2}D^{2}=r$, one can solve the differential equations to obtain the following D3-brane solution
\bea\nonumber
ds^{2} &=& \left(1-\f{q}{r}\right)^{1/2} ds_{L_{4}}^{2}+ \f{1}{r^{6}}\left(1-\f{q}{r}\right)^{-13/2}dr^{2} + \f{1}{r}\left(1-\f{q}{r}\right)^{-3/2} ds_{M_{4}}^{2}\\
 & + & r^{2}\left(1-\f{q}{r}\right)^{3/2}(d\tau - {\cal A})^{2},\\
F_{\mu\nu\rho\sigma r} & = & qr \left(1-\f{q}{r}\right)^{9/4}\e_{\mu\nu\rho\sigma}.
\eea
It is clear that, $L_{4}$ corresponds to the world-volume of the D3-brane. Asymptotically, as $r\to\infty$, the solution becomes $L_{4}\times M_{C}$ which can be seen by a coordinate change $d\tilde{r}\sim r^{-3}dr$, near infinity. The solution is half supersymmetry preserving state of the vacuum  $L_{4}\times M_{C}$, since the presence of D3-brane brakes half of the available supersymmetries. There is also  a horizon located at $r=q$. We finally note the non-orthogonal decomposition of the transverse directions due to the U(1) bundle structure.\\

In this paper we do not attempt to determine singularity structures of these solutions. Due to the presence of unbroken supersymmetries we believe that the solutions are stable. Another important open problem is to find realizations of unbroken supersymmetries on the solutions, which enable one to determine how Bogomolny bounds are saturated.

\section{Conclusions}

\hs{4}The brane solutions of supergravity theories have played a crucial role in recent developments in the non-perturbative string/M theories. With this experience, it is reasonable to claim that finding new solutions will also be important for future developments. In this paper, we have constructed new classes of solutions in $D=11$ and $D=10$ dimensions. The solutions obtained in sections 3 and 4 have a very similar structure with the well known solutions and can be viewed as the generalizations of them. Solutions obtained in section 6 belong to a different class in which the transverse directions do not have a single warping factor. In section 5, we have presented a general argument which allows one to construct new solutions from the old ones by replacing certain directions with more general manifolds.\\

In constructing new solutions, we have mainly used the theorem proved in \cite{ak}. The results of the present paper show that the theorem reviewed in section 2 is an important way of obtaining new supersymmetric solutions. Here, we would like to mention two open problems which can possibly be attacked by using the theorem; the first one is to find non-static cases which may lead to time dependent or stationary supersymmetric  solutions, and the second one is to construct explicit examples of non-trivially embedded brane solutions, as formally discussed in \cite{man}. \\

By the well known solution generating techniques, like applying S or T-dualities, or by dimensional reduction, one can obtain more new solutions in $D=10$ or in lower dimensions. As an interesting application of this, we note that it is possible to untwist a U(1) bundle by a T-duality transformation along the coordinate parametrizing U(1) fiber.\\

As for the usual brane solutions, one may try to identify low energy theories defined on the world-volumes. The number and the supermultiplet structure of collective coordinates depend on the realizations of unbroken supersymmetries, the singularities of the solutions and the isometries of the internal manifolds. A classification  of spaces having Killing spinors in diverse dimensions is the key ingredient for such an analysis. After identifying the low energy field theories, the next important question is to learn how to take decoupling or near horizon limits. We note that, brane solutions having transverse Einstein spaces give rise to generalizations of the original AdS/CFT dualities which correspond to compactifications on arbitrary Einstein manifolds. By defining sensible decoupling limits for other type of solutions found in this paper, one may discover interesting realizations of holographic principle.

%%%%%%%%%%%%%%%%%%%%%%%%%%%%%%%%%%%%%%%%%
%%%%%%%%%%%%%%%%%%%%%%%%%%%%%%%%%%%%%%%%%

%\pagebreak

\end{document}